\documentclass[12pt]{iopart}
\usepackage{graphicx}

\begin{document}

\title[Bright and dark Autler-Townes Rydberg  spectroscopy]{Bright  and dark Autler-Townes states in the atomic Rydberg multilevel spectroscopy}

\author{Giuseppe Bevilacqua$^{1}$, and Ennio Arimondo$^{2,3}$}

\address{
$^{1}$  Dept. of  Physical Sciences,  Earth and  Environment -  DSFTA,
University of Siena -- Via Roma 56, 53100 Siena, Italy\\ 
$^{2}$ Dipartimento di Fisica "E. Fermi", Universit\`a di Pisa, Largo B. Pontecorvo 3, 56122 Pisa, Italy\\
$^{3}$ Istituto Nazionale di Ottica-CNR, Via G. Moruzzi 1, 56124 Pisa, Italy}

\vspace{10pt}
\begin{indented}
\item[]March 2022
\end{indented}

\begin{abstract}
We investigated the Autler-Townes splitting produced by microwave transitions between atomic Rydberg states explored by optical spectroscopy from the ground electronic state. The laser-atom Hamiltonian describing the double irradiation of such a multilevel system is analysed on the basis of the Morris-Shore transformation. The application of this transformation to the microwave-dressed atomic system allows the identification of bright, dark, and spectator states associated with different configurations of atomic states and microwave polarisation. We derived synthetic spectra that show the main features of Rydberg spectroscopy. Complex Autler-Townes spectra are obtained in a regime of strong microwave dressing, where a hybridisation of the Rydberg fine structure states is produced by the driving.
\end{abstract}

%
\vspace{2pc}
\submitto{\JPB}
%
%
%

\section{Introduction}
Coherent superposition of quantum states is an intrinsic aspect of quantum mechanics. For instance, the description of chemical molecular bonds, which represent an important step in the initial quantum mechanics development, is based on these superpositions. However, in the initial experimental spectroscopic investigation of atoms and molecules, these superpositions were not controlled. The first step in their manipulation was associated with optical pumping, where for atoms with very long relaxation times, the broadband light sources were able to produce coherent superposition, denoted as coherences. The development of narrowband tunable lasers has represented a powerful tool for the generation of coherences in a large variety of quantum systems, often with very long survival times.\\
\indent Within the construction of quantum mechanics, the coherent superposition presence leads to quantum interference, destructive or constructive. In 1976, quite unexpected and totally independently, interferences appeared in the hands of three different atomic physics researchers. Let us mention that  the contribution of Bruce Shore at Lawrence Livermore~\cite{Shore1976} is nicely described in~\cite{ShoreDavis2001}. While studying the photoionisation of atoms driven by multistep excitation, it turned out that the excitation process was blocked. As demonstrated by Shore, this block is evidence of destructive interference associated with the coherent excitation of the multilevel system. The second observation was made in Pisa in an experiment on sodium atoms excited by a multimode laser~\cite{AlzettaOrriols1976}.  For the magic resonant condition produced by an inhomogeneous magnetic field, the sodium atoms were pumped into a coherent superposition immune to light excitation. The experimental evidence was a dark spot at a magic point within the spatially distributed sodium fluorescence. The formation of a dark coherent atomic superposition of sodium ground-state eigenstates was macroscopically visible in the laboratory. In parallel with the experimental observation, the observed dark feature was theoretically linked to a coherent superposition of the ground state in a three-level lambda scheme~\cite{ArimondoOrriols1976}. Dark state features also appeared in a contemporary theoretical investigation of a three-level cascade system by Whitley and Stroud at Rochester University~\cite{WhitleyStroud1976}. Two years later, in a sodium experiment, the same research group~\cite{GrayStroud1978} monitored the laser frequency control of dark-state preparation. These authors named coherent population trapping the creation of a three-level interfering coherent superposition.\\
\indent Today, the preparation of dark and bright states has attracted   wide  interest in quantum control investigations. In parallel, the framework required for the preparation of the dark and bright states is precisely modelled. In 1983, Morris and Shore~\cite{MorrisShore1983} formalised the laser-atom interactions required to prepare bright and dark states, as well as spectator ones in their definition by introducing the so-called Morris-Shore (MS) transformation, which was reviewed in ~\cite{Shore2014}.\\
\indent A three-level system presents an interesting independent feature, the Autler-Townes (AT) splitting.  When one transition is driven by a intense laser, the second transition spectrum is composed by a resonant doublet with the separation between the two components determined by the laser electric field. This process, well characterised in atomic/molecular and solid state spectroscopy, has received recently a new interest within  a  different  context: the  precise  determination  of  a  microwave  (mw)  field amplitude for calibration purposes as in~\cite{PiotrowiczYakshina2011,SedlacekShaffer2012,SedlacekShaffer2013,SimonsHolloway2016,HollowayRaithel2017,SongQu2018,JingJia2020,MeyerKunz2020,PengWang2020,StelmashenkoOvsyannikov2020,Chopinaud2021,JiaZhong2021,SimonsWalker2021,CaiLiu2021,ArchimiMorsch2021}.  The mw radiation is applied to cold atoms in Rydberg states where the electric dipole moment is very large, such that even a weak mw field produces AT splitting.  Most experiments have applied linear polarisation to simplify the atomic description.  However, the Rydberg states of interest have a Zeeman structure; therefore, for generic polarisation, a multilevel system is involved in the global AT response.\\
\indent This work analyses the Rydberg mw AT splitting on the basis of the MS transformation. In a three-level system the states contributing to AT splitting represent the bright states of the absorption spectrum. Both bright and dark states appear within a multilevel Rydberg structure. The MS transformation allowed us to determine the connection between the applied laser/mw geometry and bright/dark AT atomic response. The standard approach for MS transformation is based on the numerical solution of time-dependent equations. In fact, this approach has been used in numerical simulations performed by several authors within the long list of Rydberg references quoted above. Instead, we developed an alternative search tool based on the dressed-atom approach~\cite{CohenTannoudji1996}. The Rydberg system of interest is based on a strong mw field driving the Rydberg levels, and a weak optical probe on a transition from the ground state to a Rydberg one. By treating the mw Rydberg excitation using the dressed-atom approach, the dressed eigenstates are composed of bright states when interrogated by optical radiation, and by dark or spectator eigenstates that are not accessible to optical interrogation.  The dressed-atom analysis does not include the presence of relaxation processes that introduce a linewidth in the AT spectral features. We derived synthetic spectra showing the positions and relative intensities of bright AT features.\\
\indent Rydberg states with different Zeeman multiplicities were explored in the experiments quoted above. We focus our attention on the simplest atomic excitation configuration,  from the Rydberg $nS_{1/2}$ state to the $nP_{1/2}$ and $nP_{3/2}$   states,  with an $n$ Rydberg quantum number in the  60-80 range. The fine structure splitting within the $P$ multiplet is smaller than the mw energy separation between the $nS$ and $nP$ states. Two different driving regimes were considered: weak and strong. The weak regime is characterised by a mw driving Rabi frequency smaller than the fine splitting of the $nP_{1/2,3/2}$  states.  Thus, the mw field drives separately the $nS_{1/2} \to nP_{1/2}$ or $nS_{1/2} \to nP_{3/2}$ transitions of the upper level Rydberg fine structure. Increasing the mw  Rabi frequency the strong driving regime is reached, where both transitions are driven simultaneously. Mw Rabi frequencies comparable to the fine structure  splitting are easily applied in Rydberg experiments. In this regime a hybridisation  is introduced by mw electric field,  as mentioned in~\cite{SedlacekShaffer2012}. The eigenstates of the atomic structure do not coincide with those of the additional mw-atom interaction Hamiltonian, which cannot be treated as perturbations. We show that this process leads to a non-linear dependence of the AT splitting on mw field, a result very bad for AT-based Rydberg calibration. From the atomic physics point of view, this strong regime is analogous to the break of the LS coupling by a magnetic field the so-called Paschen-Back regime~\cite{Sobelman1996}, or by an electric field, as investigated for Rydberg atoms in~\cite{RyabtsevTretyakov2002}. Such competition between different bases also appears in the creation of a dark state in a three-level $\Lambda$ scheme and its destruction by an applied magnetic field producing an energy separation of the two lower levels~\cite{BerkelandBoshier2002}. In these cases, the atomic basis is broken by a Hamiltonian acting on the same subspace; in our case, the break originates from a Hamiltonian, the mw one, connecting the first structure subspace to a separate subspace.\\
\indent This work is organised as follows. Section II discusses the basics of the MS transformation. Section III introduces three essential elements of our analysis:  the Rydberg atomic structure, dressed-atom
 description, and Rydberg detection, monitored either on the optical absorption or on the selective electric field ionisation. Section IV examines the response of several Rydberg multiple-level systems to resonant mw excitation and reports spectra obtained under different excitation regimes.

\section{Morris-Shore transformation}
The new features of the three-level and multilevel systems, whose description is more complex than that of a spin 1/2 system, stimulated the search for a natural basis, that is, a basis where the theoretical description is simplified. Important contributions to this topic have been presented in~\cite{Nienhuis1996,MilnerPrior1998,Prudnikov2004}. In this direction, the existence of dark and bright states for a system with multilevel coupling has been formalised by Morris-Shore decomposition, an example of singular-value decomposition, as stated in~\cite{Shore2014}. This transformation is applied to a system composed of two sets of degenerate states, $g$ ground and $e$ excited, where by suitable partitioning and ordering of the quantum states, the Hamiltonian has the following structure:
 \begin{equation}
     \textup{H}=
    \left( \begin{array}{cc}
        \omega_gI_g &. V \\
        V^{\dagger}& \omega_{e} I_{e} \\
      \end{array} \right).
\label{eq:MSHamiltonian}
\end{equation}
Here $I_g$ and $I_{e}$ are the square unitary matrices of dimensions $N_g$ and $N_e$, respectively, with $N=N_e+N_g$. $V$ is a rectangular matrix, of dimensions $N_g \times N_e$, and $V^\dagger$ is its Hermitian conjugate of dimensions $N_e \times N_g$. The $g$ states are degenerate with energies $\omega_g$, and also the $e$ states with  $\omega_e$ energies, at $\hbar=1$. The  transitions are not allowed within these sets of states. The MS transformation replaces the $N$-linked states with a set of $N_C $ independent two-state bright (coupled) systems. The remaining uncoupled states, $N_U$  in number, are unpaired and unaffected by the $V$ and $V^\dagger$ interactions. Their number is given by
\begin{equation}
N_U=|N_g-N_e|.
\end{equation}
If the uncoupled components are in the $g$ set, that is, they are not linked to excited states, they are dark states. If the components of the unpaired state are in the $e$ set, they are denoted as spectator states by Morris and Shore. A generalisation of this transformation to the case where the blocks of the ground or excited levels are not degenerate was presented in a recent publication~\cite{ZlatanovVitanv2020}.

\section{Level scheme, dressed atom and Rydberg detection }
\begin{figure}
\centering 
  \includegraphics[angle=0, width= 0.6\columnwidth] {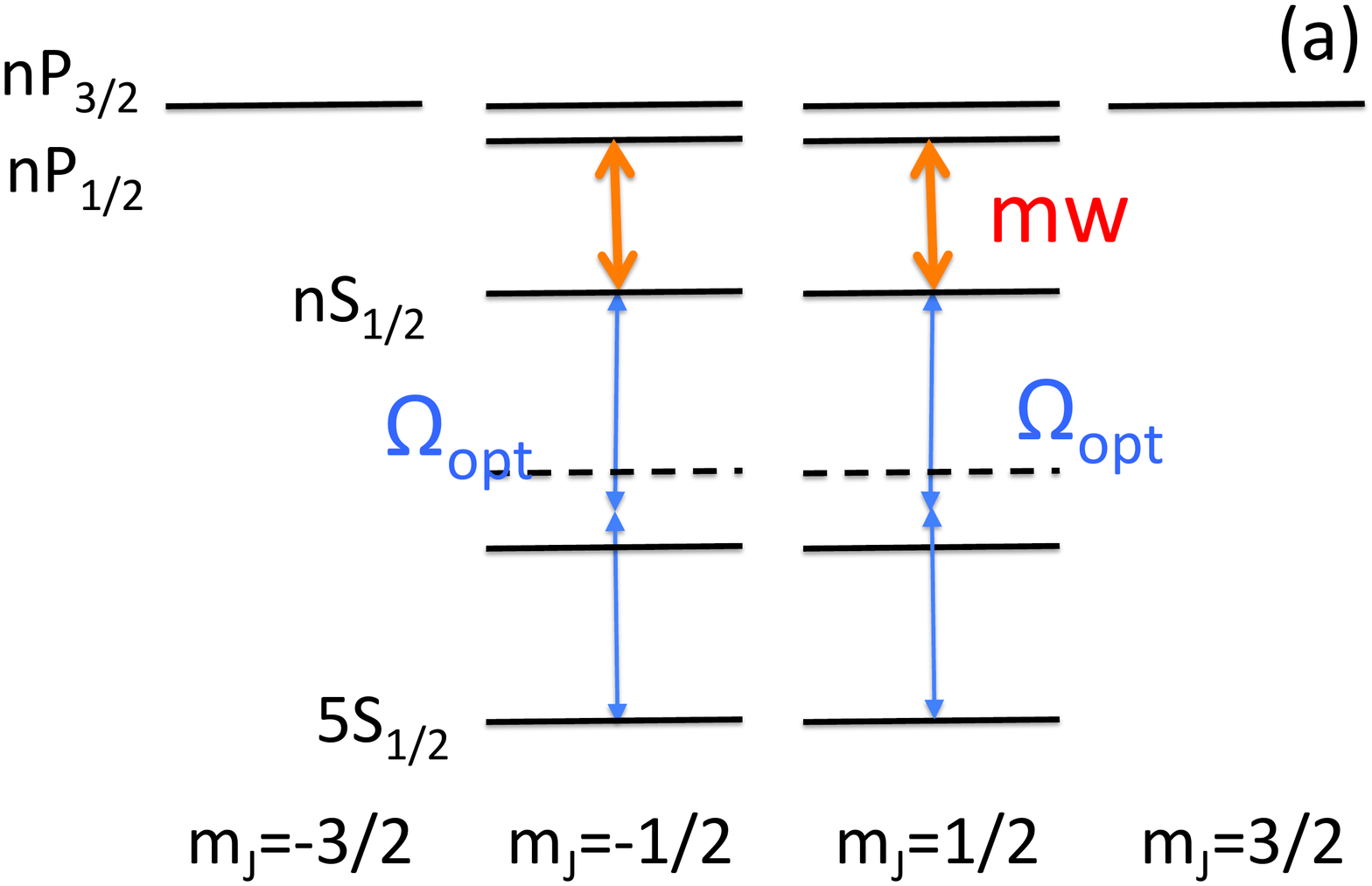}
  \includegraphics[angle=0, width= 0.6\columnwidth] {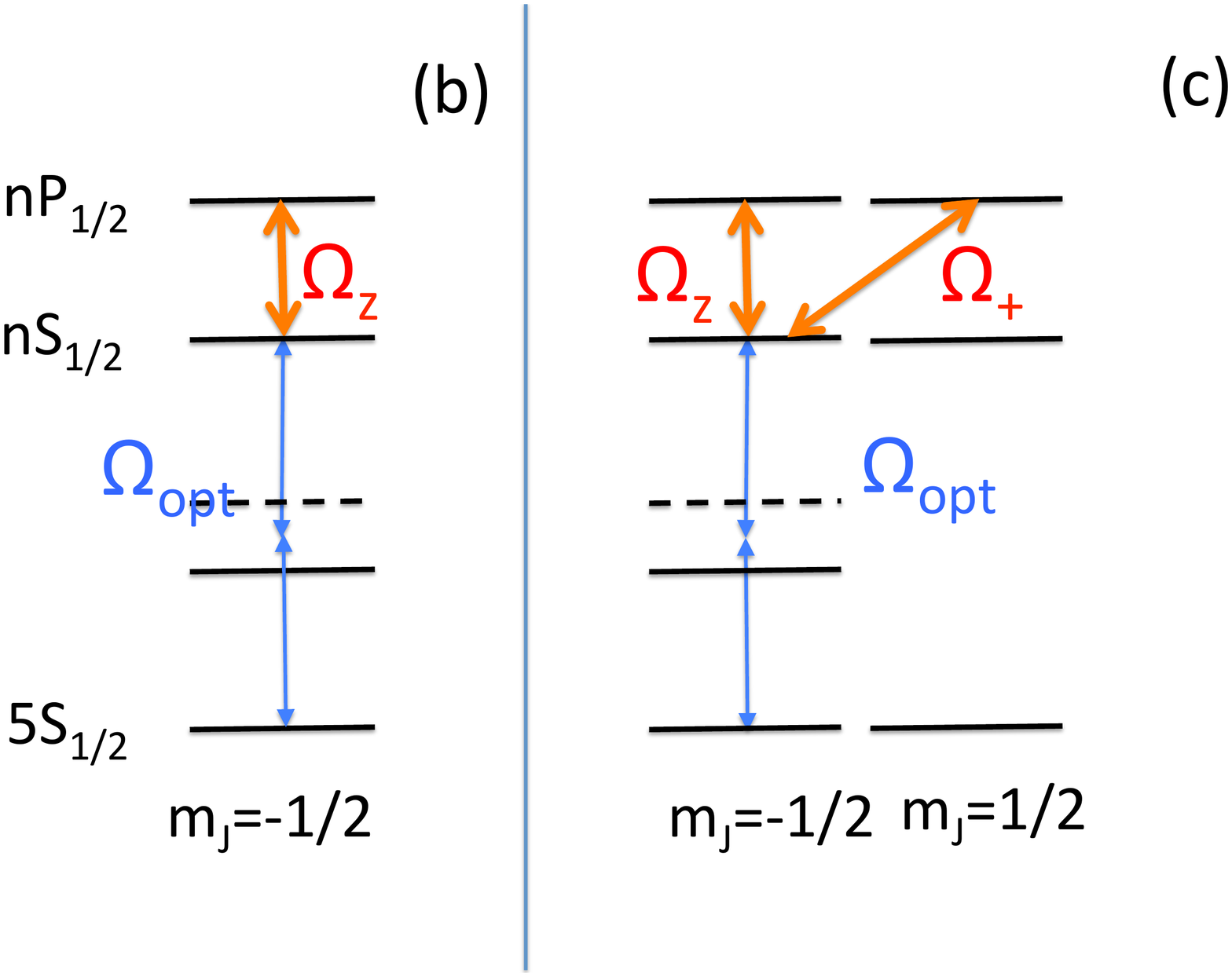}
      \caption{In (a) Schematic diagram of the $^{87}$Rb 5$S_{1/2}$  ground level and the Rydberg $nS$ and $nP$ excited ones,  with all their Zeeman structures. The Zeeman states are aligned vertically for a given $m_J$ quantum number The thin blue vertical line denotes the two-photon excitation of the $nS_{1/2}$ Rydberg level by a weak $\pi$ polarised radiation with $\Omega_{opt}$ Rabi frequency. The thick red  lines connecting the Rydberg states represent the strong $\pi$ polarized mw excitation. Mw excitation of both $nP_{1/2}$ or $nP_{3/2}$ states should be considered for mw Rabi frequencies comparable to the P state fine structure splitting. In (b) a three level scheme is obtained by a selective optical excitation and z-polarised mw radiation with $\Omega_z$ Rabi frequency. In (c) a four level  scheme is driven by the selective optical excitation and mw radiations with $\Omega_x,\Omega_+$ orthogonal polarisations.   }
\label{fig:level}       
\end{figure}

The atomic-level scheme, presented in Fig.~\ref{fig:level}(a), starts with atoms  at  the  $5S_{1/2}$ ground level, as in the $^{87}$Rb experiments of Ref.~\cite{SimonsHolloway2016}. The atoms were excited by an optical two-photon transition to the $nS_{1/2}$ Rydberg level, with $n=68$ in that experiment. This optical transition is characterised  by  $\delta_{opt}$  two-photon  detuning  and  the $\Omega_{opt}$ effective Rabi frequency. The Rydberg excited atoms are transferred by mw radiation to a level close in energy; for instance, $nP_{1/2}$ or $nP_{3/2}$,  again $n=68$.  The mw transition is characterised by the $\delta_{mw}$   detuning and the electric field applied along the different spatial axes.  The optical excitation is treated theoretically as a weak perturbation  compared to the strong mw field interaction. In the AT experiments conducted in Refs. ~\cite{PiotrowiczYakshina2011,SedlacekShaffer2012,SedlacekShaffer2013,SimonsHolloway2016,HollowayRaithel2017,SongQu2018,JingJia2020,MeyerKunz2020,PengWang2020,StelmashenkoOvsyannikov2020,Chopinaud2021,JiaZhong2021,SimonsWalker2021,CaiLiu2021,ArchimiMorsch2021} the detection of Rydberg atoms by electromagnetically induced transparency (EIT) optical spectroscopy or selective electric field	ionisation allows the experimentalists to probe the frequency and amplitude of the mw driving radiation. The mw field amplitude is derived from the AT modification of the Rydberg excitation. Different atomic schemes with alternative initial and final states are equivalent to the present scheme.\\
\indent For a simple three-level AT analysis, we consider a mw $z$-axis polarisation,
 combined with the initial atomic occupation of the ground $|5S_{1/2},m_J=-1/2\rangle$ state. After optical two-photon excitation  to  the  $|nS_{1/2},m_J=-1/2\rangle$ state,  the  atoms  are  transferred by mw radiation to the $|nP_{1/2},m_J=-1/2\rangle$ state,  as denoted by blue and red arrows in Fig.~\ref{fig:level}(b). Following the dressed-atom AT treatment described in Ref.~\cite{CohenTannoudji1996}, the quantised mw field  can be described by its $N$  photon number.   The dressed states $|nS_{1/2},m_J=-1/2;N+1\rangle$ and $|nP_{1/2},m_J=-1/2;N\rangle$ are nearly resonant and coupled by the $\Omega_z$ mw Rabi frequency.  At the $\delta_{mw}=0$ resonance, they are degenerate. Their mw coupling determines the dressed eigenstate $|1(N)\rangle$ and $|2(N)\rangle$ linear combinations of the nearly resonant states~\cite{CohenTannoudji1996}. Because both the $|1(N)\rangle$ and $|2(N)\rangle$ dressed states contain an admixture of the $nS_{1/2}$ state, the optical absorption from the ground $5S_{1/2}$ level is composed by the AT doublet.  The two AT absorption lines are separated in frequency by the $\sqrt{\delta_{mw}^2+\Omega_z^2}$  splitting of dressed states. From the viewpoint of the dark/bright states, for this $\pi$ polarised mw radiation, all dressed states interact with the optical radiation and therefore are bright states. For such $z$-axis polarisation the ground $m_J=1/2$ state  has an identical AT response.\\ 
\indent Fig. 1(c) presents a different mw driving configuration based on two orthogonal mw polarisations exciting the Rydberg $|nS_{1/2},m_J=-1/2\rangle$ initial state to both  $|nP_{1/2},m_J=\pm 1/2\rangle$ Zeeman states.  We may apply
to this system, the MS transformation, or the equivalent transformation to a "natural" basis composed by a "coupled" linear combination of $|nP_{1/2},m_J=-1/2\rangle$ and $|nP_{1/2},m_J=1/2\rangle$ eigenstates and their orthogonal "uncoupled" combination.  Those
transformations reduce the Rydberg system to two bright states and one spectator state (dark-excited state). Even in this case the AT spectrum is composed by a doublet.\\
\indent In most experimental investigations, mw absorption is detected  by  the modification of the optical EIT produced by mw driving. The EIT signal may be derived from atomic susceptibility, as in Ref.~\cite{HollowayRaithel2017}. We do not investigate this detection, which requires a numerical solution of the density matrix equations. Instead, we focus on detection by optical absorption or by the ionisation of the Rydberg final  states, as
derived from the dressed-atom approach. The mw-dressed state admixtures contain both $|nS\rangle$ and $|nP\rangle$ states. The $nS$ admixture determines the strength of the optical excitation from the ground state and reproduces the spectral features of the absorption spectrum. The Rydberg selective ionisation monitors the $nP$  admixture of each dressed state The relative strength of each resonance in the spectrum of the detected ions is derived from the product of the $nS$ admixture, determining the absorption strength, and the $nP$ admixture, determining the ionisation probability.\\
\indent In our numerical analysis not applied to a specific atomic configuration, we consider the case of fine structure splitting with round number $300$ MHz for  the $nP$   levels
corresponding to $n \approx 70$ Rb state, as in Refs.~\cite{LiGallagher2003,SibalicAdamsWeatherhill2017}. For
these states, the hyperfine splittings are negligible with no role played by the atomic nucleus. Mw Rabi frequencies are determined by transition electric dipole moments and electric field amplitudes. The dipole moments can be derived by combining for instance  the data  from~\cite{SimonsHolloway2016} and the definitions of~\cite{SibalicAdamsWeatherhill2017}. The theory presented in the following relies on the ratio between Rabi frequency and fine structure splitting. In the experimental investigation by Chopinaud and Pritchard~\cite{Chopinaud2021} mw Rabi frequencies with values up to 300 MHz were applied to caesium Rydberg states, and this range is explored in our analysis.
 
 \section{Multilevel system driving}
 To deal with several MW couplings within the structure shown in Fig. ~\ref{fig:level}(a), the entire level system should be considered. Our Rydberg atomic basis is given by $L,J,m_J$ quantum numbers.  $|L_J,m_J;N\rangle$ dressed atomic states, denoted by the $L$ symbol, the $J,m_J$  values and the $N$ mw photon number are
 \begin{eqnarray}
&|S_{1/2},1/2;N+1\rangle,|S_{1/2},-1/2;N+1\rangle,\nonumber \\
&|P_{1/2},1/2;N\rangle,|P_{1/2},-1/2;N\rangle,
\end{eqnarray}
for the $nS_{1/2} \to nP_{1/2}$ transition, and
\begin{eqnarray}
&| S_{1/2},1/2;N+1\rangle,|S_{1/2},-1/2;N+1\rangle,\nonumber \\
&|P_{3/2},3/2;N\rangle,|P_{3/2},1/2;N\rangle, |P_{3/2},-1/2;N\rangle,&|P_{3/2},-3/2;N\rangle.
\end{eqnarray}
for the $nS_{1/2} \to nP_{3/2}$ transition.\\
\indent The mw-atom coupling is determined by the electric dipole moment between the initial and final states and by the applied electric field amplitude. According to~\cite{SibalicAdamsWeatherhill2017}, the dipole
moment proportional to $\langle n;L||eR|| n';L'\rangle$ reduced dipole moment is given by:
\begin{eqnarray}
    & \langle n;L_J, m_J|er_q|n';L'_{J'},m'_{J'}\rangle= (-1)^{J-m_J+S+L+J'+1}\nonumber \\
    &\left( \begin{array}{ccc}
        J&1&J'    \\
-m_J&q&m'_{J'}\\
\end{array} \right)
      \left( \begin{array}{ccc}
        J&1&J'    \\
        L'&S&L\\
      \end{array} \right)\sqrt{(2J+1)(2J'+1)}\langle n; L||eR||n'; L'\rangle,
\end{eqnarray}
where  $er_q$ is the $q$-th spherical component of the electron dipole vector, $L$ the angular momentum, $S$ the atomic spin, and the brackets and curly brackets are the Wigner 3-j and 6-j symbols, respectively. An important feature is associated with the levels shown in Figure 1(a). $|n;S_{1/2}, m_J=\pm1/2\rangle \to |n; P_{3/2}, m_J=\pm3/2\rangle $  transitions are closed, with unitary
oscillatory strength. Therefore, within a spectroscopic approach, the dipole moments are scaled to their extreme values, that is, a renormalisation is applied to the above reduced dipole moment, as presented in the equations in~\ref{Appendix}.\\
\indent Our AT analysis deals with the case of a mw electric field having components $\left(E_x \cos(\omega_{mw} t), 0,E_z\cos(\omega_{mw} t)\right)$  corresponding to standard experimental configurations. We treat also the case of two $E_{\pm}$ electric fields, $\sigma^+$ and $\sigma^-$ polarised,  rotating and
 counter-rotating in the $(x, y)$ plane. This configuration, which has no standard for experimental investigations so far, produces interesting MS transformation features. For an $x$ polarised mw field, the $\sigma^{\pm}$ rotating/counterrotating electric field    components are given by $E_+=E_-=E_x/2$.  Elliptical field  in the $(x,y)$ plane produces different values for these components. \\
\indent While the standard Rabi frequency definition includes the transition dipole, in order to deal with the multiple transitions driven by the polarised mw electric field, we introduce the following non-standard definition of the Rabi frequencies:
\begin{equation}
\Omega_{i}=\langle n, 0||eR||n',1 \rangle_{eff} E_i,
\end{equation}
where $i=(x,z,+,-)$, and the effective dipole moment introduced in Eqs.~\ref{eq:reduceddipole}. The numerical coefficient associated with each dipole transition reported in these equations is not included in the Rabi frequencies. Instead, it will appear within the Hamiltonian definition. This approach is more convenient for comparison with the experiments, where the control is on the amplitude of the electric field components. A similar approach was applied in ref.~\cite{KisVitanov2003}. Owing to the above Rabi frequency definition, the rotating wave approximation should
be applied to the $(x,z)$ components, but is already included in the $(+,-)$ components.\\
\indent Two separate AT treatments are required when the mw Rabi frequencies are smaller than the upper state fine structure splitting (weak driving regime) and when these frequencies are comparable to that splitting (strong driving regime). The AT spectra in the following were calculated by assuming $\pi$ optical excitation for the equally populated S ground states.

\subsection{nS$_{1/2}$-nP$_{1/2}$ driving}  
\indent For this case and the different mw polarisations, the 4 $\times$ 4 dressed-atom Hamiltonian is cast in the following form of Eq.~\ref{eq:MSHamiltonian}:
\begin{equation}
\omega_{S}=0, \qquad
\omega_{P_{1/2}}=-\delta^{1/2}_{mw},
\end{equation}
\begin{equation}
\label{eq:P12} 
    V_{x,z}^{P12}=\frac{1}{2}
     \left( \begin{array}{cc}
\frac{1}{\sqrt{3}}\Omega_{z}  &\sqrt{\frac{2}{3}}\Omega_{x}\\
\sqrt{\frac{2}{3}}\Omega_{x}  & -\frac{1}{\sqrt{3}}\Omega_{z}\\
\end{array} \right),  \qquad
    V_{+,-,z}^{P12}=
    \left( \begin{array}{cc}
\frac{1}{2\sqrt{3}}\Omega_{z}  &\sqrt{\frac{2}{3}}\Omega_{-}   \\
\sqrt{\frac{2}{3}}\Omega_{+}  & - \frac{1}{2\sqrt{3}}\Omega_{z}\\
\end{array} \right). 
\end{equation}
with $\delta^{1/2}_{mw}$ the mw detuning for $S\to P_{1/2}$ driven Rydberg transitions.\\  
\indent For the majority of Rabi frequency values, four separate eigenstates exist. From the dressed eigenstates, we derive that all four are bright for the probe.
 optical transition. For the $(+,-,z)$ basis at $\delta^{1/2}_{mw}=0$ the four dressed eigenvalues are given by
 \begin{equation}
\lambda=\pm\frac{1}{\sqrt{3}}\left[ 
\Omega_z^2/4+2\Omega_+^2+2\Omega_-^2 \pm \sqrt{2}\left(\Omega_+-\Omega_-\right)\sqrt{SR}\right]^{1/2},
\label{eq:sp12_pmz}
\end{equation}
with 
\begin{equation}
SR=2\left(\Omega_++\Omega_-\right)^2+\Omega_z^2.
\end{equation}
However, the four eigenvalues may collapse into two degenerate ones, and the AT spectrum reduces to the standard spectrum for two degenerate three-level systems. This applies to the $\Omega_x,\Omega_z$  configuration with double-degenerate values given by
\begin{equation}
\lambda_\pm=\frac{-\delta_{mw}^{1/2}\pm\sqrt{(\delta_{mw}^{1/2})^2+\frac{1}{3}(\Omega_x^2+2\Omega_z^2)}}{2}.
\label{eqsp12xz}
\end{equation}
An equivalent AT eigenvalue equation is reported in~\cite{Chopinaud2021} , except for the Clebsh-Gordan coefficients. From Eq.~\ref{eq:sp12_pmz}, we derive that degeneracy also occurs on the $\Omega_+=\Omega_-$ bisector for all $\Omega_x$ values. For the $\Omega_+$ and $\Omega_-$ axes in the $\Omega_z=0$ plane, only two bright states appear combined with a dark state and a spectator one. For $\delta^{1/2}_{mw}=0$ Eq.~\ref{eqsp12xz} shows a dependence of the AT shifts on the $(x,z)$ components described by a proportionality to the mw field modulus,  the components being weighted by the Clebsh-Gordan coefficients. For $(+,-,z)$ basis Eq.~\ref{eq:sp12_pmz} shows a more complex dependence of the AT shifts on mw driving strength.\\
\begin{figure}
\centering 
  \includegraphics[angle=0, width= 0.6\columnwidth] {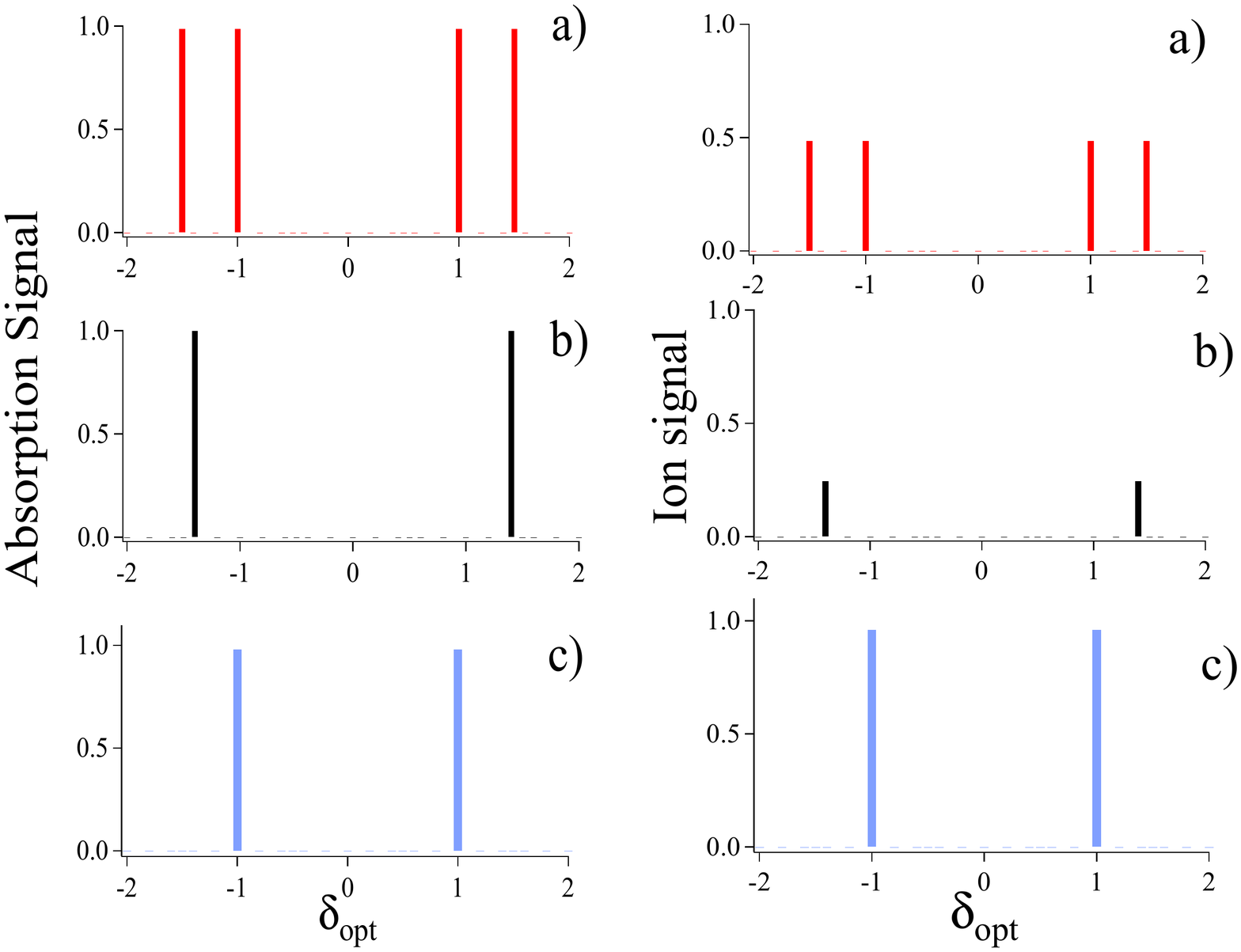}
      \caption{Synthetic spectra with positions and intensities of the $nS_{1/2} \to nP_{1/2}$ AT resonance peaks detected in the optical probe absorption (on the left column) and in the ion signal (on the right column). Both absorption and ion signals are in relative values. In a) four AT peaks appear for $(\Omega_z=2\sqrt{3}, \Omega_+=\sqrt{3/2}, \Omega_-=\Omega_+/2)$. The four peaks collapse in two ones, in b) for $(\Omega_z=2\sqrt{3},\Omega_\pm=\sqrt{6})$, and in c) for $(\Omega_z=2\sqrt{3},\Omega_\pm=0)$. Resonance positions  derived by the dressed eigenvalues, and strengths from the eigenstates as in the text.}
\label{fig:ATSP12}       
\end{figure} 
\indent Fig. 2 shows a schematic view of the frequency positions and intensities of the AT resonances. The left column plots show the optical absorption for different  values of the Rabi frequencies as derived from the admixture of the dressed states by assuming equal initial populations in the $5S$ ground Zeeman levels. The right column plots the ionisation signal determined by both the $nS$ admixture and the occupation of the $nP$ ionised Rydberg state. The AT spectra for absorption and ion detection are similar. The relative peak heights depend on the $nS$ and $nP$ components of the dressed eigenvectors. The spectral features depend only on the ratio of the Rabi frequencies. For a given ratio, the AT resonance positions scale with Rabi amplitudes. Plots (a) of the figure show a four-bright-state configuration. For the driving parameters of (b) and similar Rabi frequency ratios, the four eigenvalues collapse into two degenerate bright ones. The AT spectrum is equivalent to that of a two-degenerate three-level system with two bright states. Such a spectrum also appears for the $(\Omega_x,\Omega_z)$ configuration corresponding to equal $\Omega_+$ and $\Omega_-$ components Also in c) plots for $\Omega_z$  only different from zero, two coincident bright states appear.\\
\indent The spectra derived by the dressed atom do not contain the relaxation processes and therefore have a zero linewidth. For optical frequency scanning, as in the experimental spectra similar to those in Fig.~\ref{fig:ATSP12}, the linewidths are in the 5 MHz range, being determined from the first step of the two-photon $5S \to nS$ transition.

 \subsection{nS$_{1/2}$-nP$_{3/2}$ driving} 
\begin{figure}
\centering 
  \includegraphics[angle=0, width= 0.5\columnwidth] {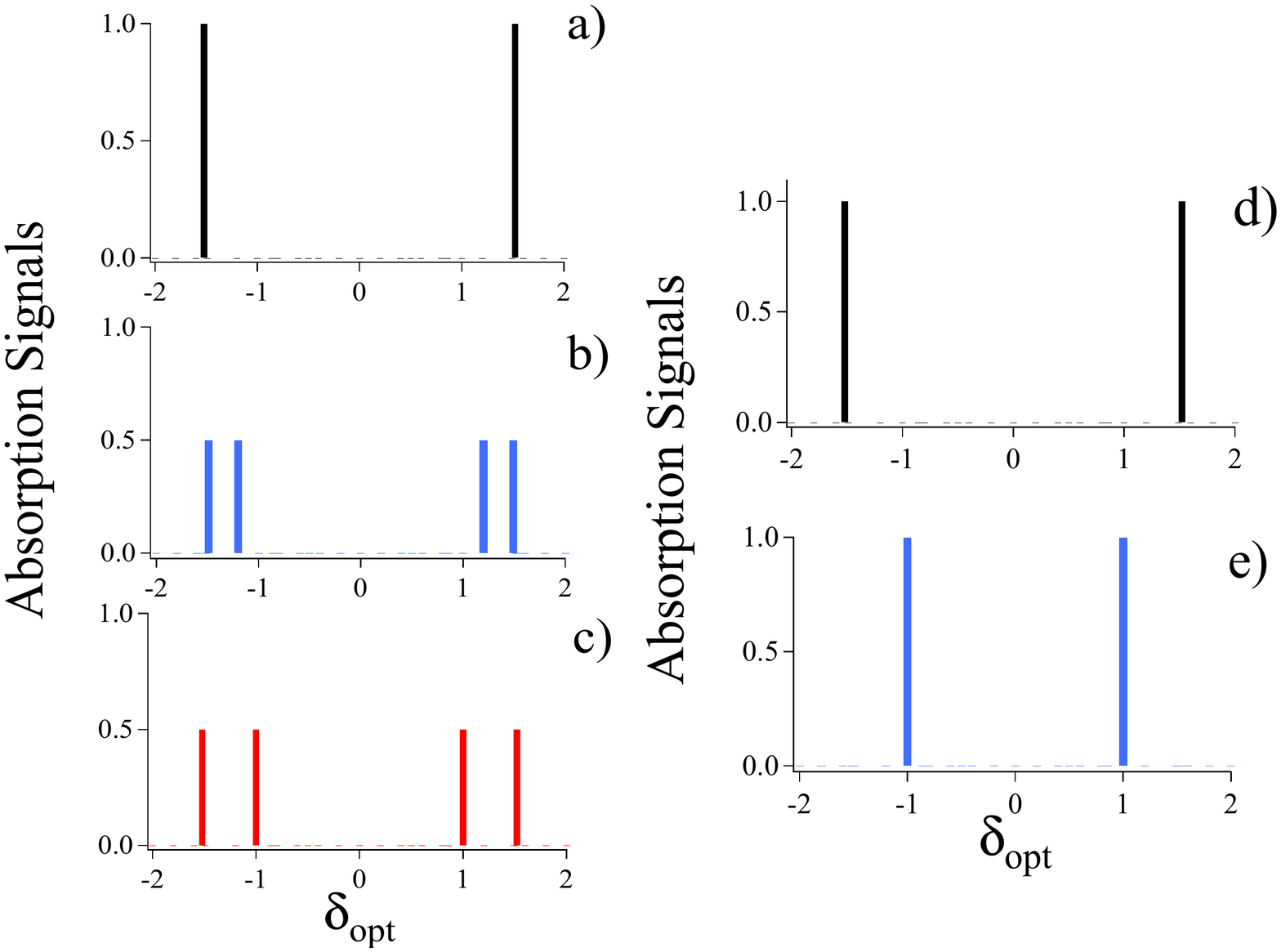}
      \caption{Synthetic spectra with positions and intensities of the $nS_{1/2} \to nP_{3/2}$ AT bright peaks detected in the optical probe absorption, measured in relative values. Rabi frequencies $(\Omega_+, \Omega_-,\Omega_z)$  in a) $(1,0,\sqrt{3/2})$, in b) $(1,0.5,\sqrt{3/2})$, and in c) $(1,1,\sqrt{3/2})$, Rabi frequencies $(\Omega_x,\Omega_z)$ $(2,\sqrt{6})$, and $(0,\sqrt{6})$  in d) and e), respectively. Four AT peaks appear in the b)-c) plots, and only two in the remaining  ones. Resonance positions  determined by the dressed eigenvalues, and strengths from the eigenstates as in the text.}
\label{fig:ATSP32}       
\end{figure}

\indent In this case the 6 $\times$ 6 Hamiltonian is  cast in the follwing  form of Eq.~\ref{eq:MSHamiltonian}:
\begin{eqnarray}
\omega_{S}&=&0,\nonumber \\
\omega_{P_{3/2}}&=&-\delta_{mw}^{3/2}.
\end{eqnarray}
\begin{eqnarray}
\label{eq:P32} 
    V_{x,z}^{P32}&=&\frac{1}{2}
     \left( \begin{array}{cccc}
\Omega_{x}  &-\sqrt{\frac{2}{3}} \Omega_{z}&-\frac{1}{\sqrt{3}}\Omega_{x} &0\\
0  &\frac{1}{\sqrt{3}}\Omega_{x}&-\sqrt{\frac{2}{3}}\Omega_{z}&-\Omega_{x }\\
\end{array} \right),  \nonumber \\
    V_{+,-,z}^{P32}&=&
    \left( \begin{array}{cccc}
\Omega_{+}  &-\frac{1}{\sqrt{6}} \Omega_{z}&-\frac{1}{\sqrt{3}}\Omega_{-} &0\\
0  &\frac{1}{\sqrt{3}}\Omega_{+}&-\frac{1}{\sqrt{6}} \Omega_{z}&-\Omega_{-}\\
\end{array} \right). 
\end{eqnarray}
\indent The AT response for this case is similar to that in the previous case, except for the presence of two spectator states associated with all Rabi frequency values. Therefore, the spectrum contains a maximum of four separate bright AT peaks, as in b) and c) plots on the left column of Fig.~\ref{fig:ATSP32}. The four eigenstates may collapse in the two degenerate ones as shown in the remaining plots of that figure. The left column spectra appear in the
$(+,-,z)$  mw field     configurations with increasing $\Omega_-$ values. The ones on the right are associated to  the (x, z) mw configuration. As for the previous $P_{1/2}$ case positions and intensities of AT peaks are derived from the dressed-atom approach. The previous discussion of spectral linewidths also applies to the present case. The ion signal spectra, not reported in the figure, were similar except for the relative intensity of the peaks.\\
 
\subsection{Strong driving}
At large intensities of mw radiation, simultaneous excitation of both $nP_{1/2,3/2}$ fine structure levels occurs. Therefore, the mw transformation should be applied to the following Hamiltonian more generally than the previous one:
\begin{equation}
  \label{eq:MSHamiltonian2}
      H=
      \left( \begin{array}{ccc}
      \omega_gI_g & V_{e1}&V_{e2}\\
      V_{e1}^{\dagger} & \omega_{e1} I_{e1}&0\\
     V_{e2}^{\dagger}&0&\omega_{e2} I_{e2}
       \end{array} \right).
 \end{equation}
 owing  to  the  presence  of  two  nondegenerate  excited  states  linked  by  a  single electromagnetic field to the ground state.  Here, $|e_1\rangle$ and $|e_2\rangle$ represent $P_{1/2}$ and $P_{3/2}$ excited states. For degenerate excited states, we recovered the Hamiltonian scheme in Eq. ~\ref{eq:MSHamiltonian}.\\
 \indent In the LS basis, the atomic variables are the electron spin and orbital momentum in both the S and P states, that is, $\vec{s}^S$ and  $\vec{s}^P$ for the spin, and $\vec{L}^P$ for the orbital momentum. At the P levels, the fine splitting is produced by the following Hamiltonian:
 \begin{equation}
H_{fs}=A_{fs}\vec{s}^P\cdot\vec{L}^P,
\end{equation}
with $A_{fs}$ the fine structure constant. The fine structure coupling is diagonalised on the $|J,m_J\rangle$ basis. On this basis, the diagonal components of the 8 $\times$ 8 dressed atom
Hamiltonian cast in the  following form of Eq.~\ref{eq:MSHamiltonian2}:
\begin{eqnarray}
\omega_{S1/2}&=&
\left( \begin{array}{cc}
0&0\\
0&0 
\end{array} \right),    \nonumber \\
\omega_{P1/2}&=&
\left( \begin{array}{cc}
-2A_{fs} -\delta_{mw}  &0\\
0 & -2 A_{fs} -\delta_{mw}\\
\end{array} \right),    \nonumber \\
 \omega_{P3/2}&=&
\left( \begin{array}{cccc}
A_{fs}-\delta_{mw}&0&0&0\\
0&A_{fs}-\delta_{mw}&0&0\\
0 &0&A_{fs}-\delta_{mw}&0\\
0&0&0&A_{fs}-\delta_{mw}\\
\end{array} \right).    
 \end{eqnarray}
where $\delta_{mw}$ represents a generic detuning for the mw-driven Rydberg transitions defined with reference to the centre of gravity of the P multiplet, that is, imposing $A_{fs}=0$.  The Hamiltonian-outHamiltonian out diagonal blocks are derived from the previous cases with
 \begin{equation}
 V_{e1}= V^{P12}; \qquad V_{e2}=V^{P32},
 \end{equation}
 including their mw polarisation dependences as in Eqs.~\ref{eq:P12}  and ~\ref{eq:P32}.  \\
\indent At low values of Rabi frequencies, which are smaller than the $A_{fs}$  splitting, the present case reduces to those examined previously. Instead, a large difference appears at large Rabi frequency values in the strong regime with hybridisation of the fine structure states introduced by the mw electric field, as mentioned in~\cite{SedlacekShaffer2012}.  The eigenstates of the atomic structure do not coincide with those of the additional laser-atom interaction Hamiltonian, which cannot be treated as a perturbation. Rabi frequencies applied
in a few Rydberg experiments are large enough to produce such an mw hybridisation. The hybridisation between hyperfine coupling and AT splitting was examined in Ref.~\cite{KirovaEkers2017} is equivalent to that in the present study.\\
\begin{figure}
\centering 
  \includegraphics[angle=0, width= 0.7\columnwidth] {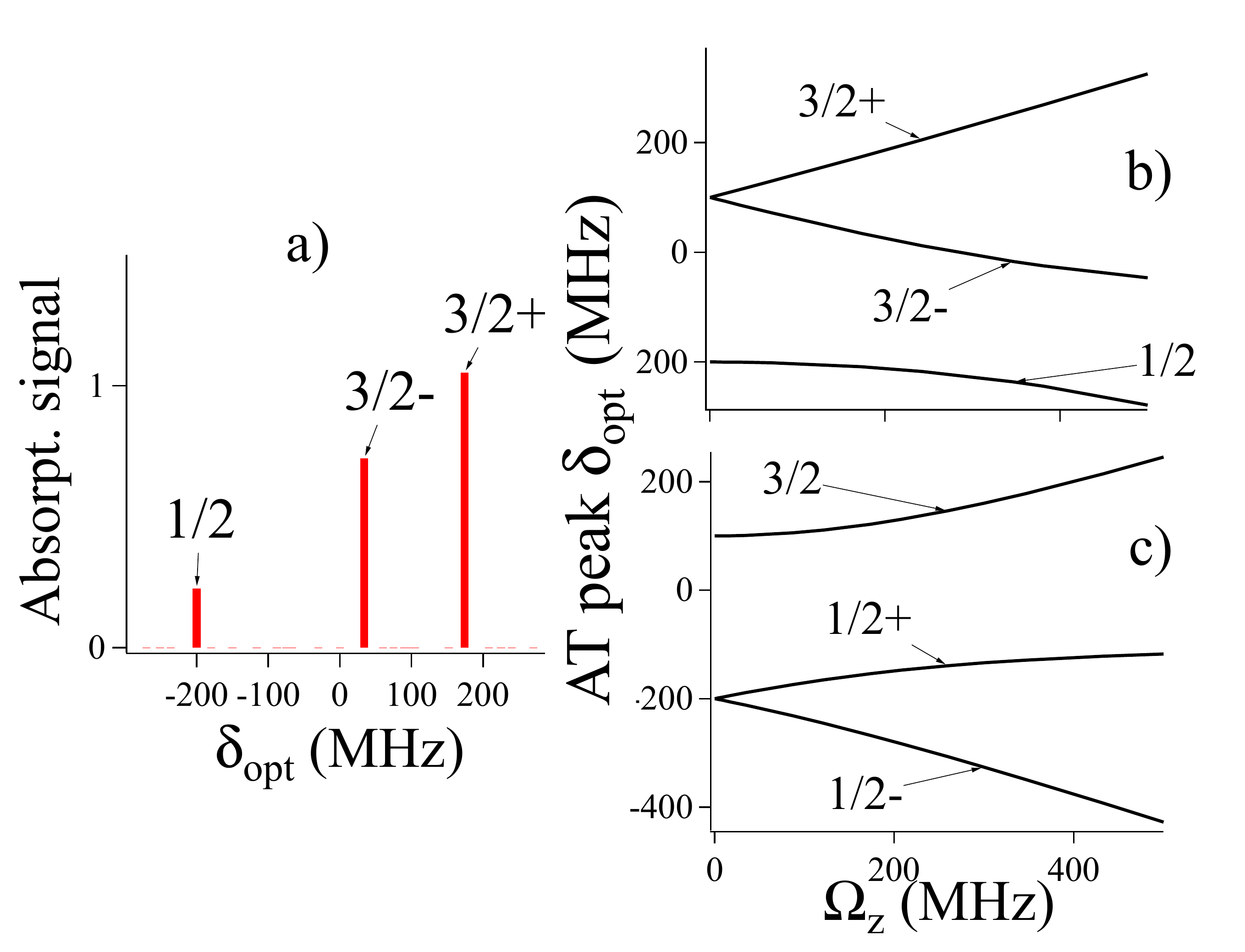}
 \caption{In (a) positions and intensities of the AT resonance peaks detected on the optical probe absorption for a $\pi$ polarised mw driving resonant with the $nS \to nP_{3/2}$ transition for a fine-structure splitting of 300 MHz and $\Omega_z=100\sqrt{3}$ MHz. 
The AT peaks, denoted as 3/2+, 3/2- and 1/2, shifted from $\delta_{opt}\approx 100$ MHz and $\delta_{opt}\approx -200$ MHz positions of resonances occurring at very small $\Omega_z$ values.  In the remaining plots, the $\delta_{opt}$ position frequencies of the three AT peaks are plotted against the applied  $\pi$ driving $\Omega_z$ value. (b) mw field resonant with
the  $nS \to nP_{3/2}$  transition, and c) resonant with  $nS \to nP_{1/2}$  transition.}
\label{fig:ATstrong}       
\end{figure} 
\indent The bright/dark states of the full Hamiltonian remain those of the previous cases with a maximum of six independent bright eigenvalues, reached for$(\Omega_-,\Omega_+,\Omega_z)$  driving with $\Omega_- \ne\Omega_+$   A minimum of two spectator states are present in all mw configurations. A new feature of a non-linear dependence of the AT splitting on the mw field  amplitude appears for all mw polarisations. Plot a) in Fig.~\ref{fig:ATstrong} shows the AT main features for a $\pi$-polarised mw field  at $\delta_{mw}$ resonant with the $nS \to nP_{3/2}$   transition upshifted by fine structure splitting.  That  AT  synthetic spectrum presents the position and intensities of three doubly-degenerate AT peaks produced by strong mw driving. The two upper frequency peaks are upshifted and downshifted with respect to the $\delta_{opt} = 100$ MHz optical absorption resonance associated with the mw not-driven case. These peaks correspond to those presented in the previous subsection for mw driving only the $nS \to nP_{3/2}$ transition. The main difference is that, in the present case, the two peaks are asymmetric in both positions, eighteen percent, and intensity, nine percent. An important new feature in the a) synthetic spectrum is the presence of a weak AT third peak produced by the off-resonance mw excitation of the $nP_{1/2}$ state. This low-frequency AT peak corresponded to the $nS \to nP_{1/2}$  transition occurring at $\delta_{opt}  = -200$ MHz for the
mw not-driven case. The difference from this value is produced by the ac Stark shift of the applied mw radiation.  Similar multipeak atomic responses were obtained for all the mw polarisations. In Ref.~\cite{ZlatanovVitanv2020}, where a generalised MS transformation is applied to Hamiltonians equivalent to Eq.~\ref{eq:MSHamiltonian2}, the triplets of states represent the extension of the bright/dark pair states of the standard MS transformation. These triplets were equivalent to the triple AT peaks shown in Fig.~\ref{fig:ATstrong}.\\
\indent The plots b) and c) of the figure show the dependence on $\Omega_z$  mw field magnitude for the frequencies of the three bright AT absorption peaks, in b) for an $nS_{1/2} \to nP_{3/2}$  resonant mw field,  and in c) for an $nS_{1/2} \to nP_{1/2}$  resonant mw field. At $\Omega_z=0$, the absorption peaks are shifted from the $\delta_{opt} = 0$ fine structure centre of gravity by 100 MHz for the first case and by -200 MHz for the second case. The most important feature of these plots is the nonlinear dependence of the AT shifts on the mw RAbi frequency, i.e., the electric field   amplitude. For the Rydberg experiments aiming to an absolute calibration of the mw electric field amplitude it is important to notice that even at Rabi frequencies less than 5 MHz, those two peaks preserving a quasi linear dependence on the Rabi frequency are not symmetric. They
had a slope difference around ten percent for $nS_{1/2} \to nP_{3/2}$ driving.. Similar results for nonlinearity and asymmetry appear in the c) plot of $nS_{1/2} \to nP_{1/2}$ driving, with resonance asymmetry reduced by a factor of two.

\subsection{nP$_{3/2}$, nP$_{1/2}$ - nS$_{1/2}$ driving}
While the previous analysis is concentrated on $nS \to nP$ mw driving, we consider here the reverse-level scheme with optical excitation to the $nP_{3/2}$  state and microwave driving in an $nS_{1/2}$ state. Such a scheme is difficult to implement in Rydberg experiments because in two-photon excitation from the $S$ ground state, the $nP$ states cannot be reached directly owing to the dipole selection rules, and single photon excitation is technically difficult to implement.  Instead, several investigations concentrated on the $nD_{3/2,5/2} \to nP_{1/2,3/2}$ configuration [~\cite{SedlacekShaffer2012,HollowayRaithel2017,JingJia2020,PengWang2020,JiaZhong2021}. All of these multilevel schemes deal with the mw driving applied between an initial state whose Zeeman multiplicity is larger than the final   one. However, from the perspective of MS transformation, the bright and dark/spectator states do not depend on the choice of the initial and final states. Therefore, the analyses of the previous subsections
for $nS \to nP{3/2}, nP_{1/2}$  driving also applies to the reverse-driving scheme.  The only differenceis that in this last scheme the spectator states become dark states.

 \section{Conclusions}
 We consider the AT splitting produced by a mw field driving a transition between the Rydberg atomic states, an important issue for the metrology of mw fields. Our analysis is based on the dressed atom description for the Rydberg states interacting with mw radiation probed by a weak optical radiationconnecting the ground atomic state to the excited states. The MS transformation was applied to. such atomic system. The eigenvalues and eigenstates of the dressed Hamiltonian represent the bright and dark spectator states of the MS transformation. These eigenvalues/eigenvectors allowed us to derive synthetic spectra for AT Rydberg spectroscopy.  This approach is also applied to a strong driving regime where a hybridisation of the structure levels is produced by mw radiation, in which the linear dependence of the AT splitting on the amplitude of mw driving is modified. This  hybridisation role is an important issue for metrology precision.  Our investigation of the AT atomic response as a function of the mw field polarisation highlights the importance of its precise control.  In addition, the AT spectra may be used to probe the different mw polarisation geometries.\\
\indent Our approach did not include relaxation processes introducing a linewidth to our synthetic spectra. For such a complete analysis, MS transformation should be applied to the density matrix equations of the driven atomic system. Our synthetic spectra were obtained for a sweep of the optical field exciting the ground atoms to a Rydberg state. For this configuration, the linewidth of each bright resonance is approximately 5 MHz in the absence of saturation broadening. Owing to the long relaxation times of the Rydberg states, narrower linewdiths are obtained by scanning mw field   at a fixed    optical excitation frequency. This approach is not practical in the metrology of an unknown mw radiation, such as blackbody radiation. Instead, it represents a powerful approach to explore the complexity of AT spectra as a function of MW polarisation.\\
\indent Ref.~\cite{SimonsWalker2021}  recently introduced an advanced spectroscopic tool based on double-dressing of the Rydberg state with a secondary mw field and an additional Rydberg state. The application of MS transformation to such a more complex atomic configuration could identify the mw schemes that are most interesting for such investigations.

\section{Acknowledgments}
This work was dedicated by one of us (EA) to the memory of Bruce Shore, whose passion for breaking a complex quantum mechanics problem into small information bits has been a constant inspiration. EA thanks Oliver Morsch for pointing out the interesting issue of AT splitting in the Rydberg states. The authors thank Teodora Kirova for verifying the dipole-moment matrix elements.

\appendix
\section{Dipole moment matrix elements}
\label{Appendix}
\indent
The dipole moment matrix elements of the tensorial components $r_0,r_\pm$ are expressed as a function of the effective reduced Rabi frequencies:
\begin{eqnarray}
\label{eq:reduceddipole}
&<n,0,\frac{1}{2}\pm\frac{1}{2}|er_0|n',1,\frac{1}{2},\pm\frac{1}{2}>=\pm\frac{1}{3}<0||eR||1>_{eff} \nonumber \\
&< n,0,\frac{1}{2}\mp\frac{1}{2}|er_{\pm 1}|n',1,\frac{1}{2}\pm\frac{1}{2}>=\frac{\sqrt{2}}{3}<0||eR||1>_{eff} \nonumber \\
&<n,0, \frac{1}{2},\pm\frac{1}{2}|er_0|n',1,\frac{3}{2}\pm\frac{1}{2}>=\frac{\sqrt{2}}{3}<0||eR||1>_{eff} \nonumber \\
&<n,0,\frac{1}{2}\pm\frac{1}{2}|er_{\mp 1}|n',1\frac{3}{2}\mp\frac{1}{2}>=\frac{1}{3}<0||eR||1>_{eff} \nonumber \\
&<n,0, \frac{1}{2},\pm\frac{1}{2}|er_{\mp 1}|n',1,\frac{3}{2},\pm\frac{3}{2}>=\frac{1}{\sqrt{3}}<0||eR||1>_{eff}.
\end{eqnarray}
\indent The above result  coincides with the branching ratio value reported in page 4 of~\cite{SimonsHolloway2016} for the same optical transition.\\


\begin{thebibliography}{}
\bibitem{Shore1976} B. W. Shore, "Inaccessible Populations in Multistep Ionization; TAMP Memo 1976", Lawrence Livermore National Laboratory: Livermore, (CA) (1976).
\bibitem{ShoreDavis2001} B. Shore, M. Johnson, K.C. Kulander, and J. Davis. "The Livermore experience: Contributions of J. H. Eberly to laser excitation theory. Opt Express. {\bf 8} 28-43. (2001) doi: 10.1364/oe.8.000028. PMID: 19417783.
\bibitem{AlzettaOrriols1976} G. Alzetta, A. Gozzini,  L.  Moi, and  G. Orriols, "An experimental Method for the Observation of R.F. Transitions and Laser beam Resonances in Oriented Na Vapour",Nuovo Cimento B {\bf 36}, 5 (1976).
\bibitem{ArimondoOrriols1976} E.Arimondo and G.Orriols, "Non-Absorbing Atomic Coherences by Two-Photon Transitions in a Three-Level Optical Pumping", Lett. Nuovo Cimento {\bf 17}, 333 (1976).
\bibitem{WhitleyStroud1976} R. M. Whitley, and C. R. Stroud, Jr., "Double optical resonance", Phys. Rev. A {\bf 14}, 1498 (1976).
\bibitem{GrayStroud1978} H. R. Gray, R. M. Whitley, and C. R. Stroud, Jr., "Coherent trapping of atomic populations", Opt. Lett. {\bf 3}, 218-20 (1978).
\bibitem{MorrisShore1983} J. R. Morris, and B.W. Shore, "Reduction of degenerate two-level excitation to independent two-state systems", Phys. Rev. A {\bf 27} 906-912 (1983).
\bibitem{Shore2014} B. W. Shore, "Two-state behavior in N-state quantum systems: The Morris-Shore transformation reviewed", J. Mod. Opt. {\bf 61}, 787-815 (2014).
\bibitem{PiotrowiczYakshina2011} M.J. Piotrowicz, C. MacCormick, A. Kowalczyk, S. Bergamini, I.I.  Beterov, and E. A. Yakshina, "Measurement of the electric dipole moments for transitions to rubidium Rydberg states via Autler-Townes splitting", NJP {\bf 13}, 093012 (2011).
\bibitem{SedlacekShaffer2012} J. A. Sedlacek, A. Schwettmann, H. Kübler, R. L\"ow, T Pfau and J. P. Shaffer,   "Microwave electrometry with Rydberg atoms in a vapour cell using bright atomic resonances",  Nature Phys. {\bf 8}, 819-824 (2012).
\bibitem{SedlacekShaffer2013} J. A. Sedlacek, A. Schwettmann, H. Ku?bler, and J. P. Shaffer, "Atom-Based Vector Microwave Electrometry Using Rubidium Rydberg Atoms in a Vapor Cell", Phys. Rev. Lett. {\bf 111}, 063001 (2013).
\bibitem{SimonsHolloway2016} M. T. Simons, J. A. Gordon, and C. L. Holloway, "Simultaneous use of Cs and Rb Rydberg atoms for dipole moment assessment and RF electric field measurements via electromagnetically induced transparency", J. Appl. Phys. {\bf 120}, 123103 (2016).
\bibitem{HollowayRaithel2017} C. L. Holloway, M. T. Simons, J. A. Gordon, A. Dienstfrey, D. A. Anderson, and G. Raithel, "Electric field metrology for SI traceability: Systematic measurement uncertainties in electromagnetically induced transparency in atomic vapor", J. Appl. Phys. {\bf 121}, 233106 (2017).
\bibitem{SongQu2018} Z. Song, W. Zhang, Q. Wu, H. Mu, X. Liu, L. Zhang, and J. Qu, "Field Distortion and Optimization of a Vapor Cell in Rydberg Atom-Based radio-frequency electric field measurement", Sensors-Basel  {\bf 18}, 3205 (2018). doi:10.3390/s18103205
\bibitem{JingJia2020} M. Jing, Y. Hu, J. Ma, H. Zhang, L. Zhang, L. Xiao, and S. Jia, "Atomic superheterodyne receiver based on microwave-dressed Rydberg spectroscopy", Nature Physics {\bf 16}, 911-5 (2020). 
\bibitem{MeyerKunz2020} D. H. Meyer, Z. A. Castillo, K. C. Cox,  and P. D. Kunz, "Assessment of Rydberg atoms for wideband electric field sensing", J. Phys. B:  At. Mol. Opt. Phys. {\bf 53}, 034001 (2020).
\bibitem{PengWang2020}Y. D. Peng, J. L. Wang, C. Li, X. Lu, Y. H. Qi,  A. H. Yang, and  J. Y. Wang, "Enhanced microwave electrometry with intracavity
anomalous dispersion in Rydberg atoms", Opt. Quant. Electronics {\bf 52}, 120 (2020). 
\bibitem{StelmashenkoOvsyannikov2020} E. F. Stelmashenko, O. A. Klezovich, V. N. Baryshev, V. A. Tishchenko, I. Yu. Blinov, V. G. Palchikov, and V. D. Ovsyannikov,  "Measuring the Electric Field Strength of Microwave Radiation at the Frequency of the Radiation Transition Between Rydberg States of Atoms $^{85}$Rb", Opt. Spectr. {\bf 128}, 1067-73 (2020).
\bibitem{Chopinaud2021} A. Chopinaud  and J.D. Pritchard, "Optimal State Choice for Rydberg-Atom Microwave Sensors", Phys. Rev. Appl. {\bf 16}, 024008 (2021).
\bibitem{JiaZhong2021} F.-D. Jia, X.-B. Liu, J. Mei, Y.-H. Yu, H.-Y. Zhang, Z.-Q. Lin, H.-Y. Dong, J. Zhang, F. Xie, and Z.-P. Zhong, "Span shift and extension of quantum microwave electrometry with Rydberg atoms dressed by an auxiliary microwave field", Phys. Rev. A {\bf 103}, 063113 (2021).
\bibitem{SimonsWalker2021} M. T. Simons, A. B.  Artusio-Glimpse, C. L.  and Holloway, E.  Imhof, S. R  Jefferts,  R.  Wyllie, B. C.  Sawyer,  and T. G. Walker, "Continuous radio-frequency electric-field detection through adjacent Rydberg resonance tuning", Phys. Rev. A {\bf 104},  032824 {2021}. 
\bibitem{CaiLiu2021} M. H. Cai, Z. S. Xu, S. H. You, and H. P. Liu, "Sensitivity improvement of Rydberg atom-based microwave sensing via electromagnetically induced transparency, ArXiv:2111.06582v1.
\bibitem{ArchimiMorsch2021} M. Archimi,  M. Ceccanti, M. Distefano, L. Di Virgilio, R. Franco, A. Greco, C. Simonelli, E. Arimondo, D. Ciampini, and O. Morsch, "Measurements of blackbody radiation-induced transition rates between high-lying S, P and D Rydberg levels", arXiv:2111.15333.
\bibitem{CohenTannoudji1996} C. Cohen-Tannoudji, "The Autler-Townes Effect Revisited" In: Chiao R.Y. (eds) Amazing Light. Springer, New York, NY. (1996).
\bibitem{Sobelman1996} I. I. Sobelman, {\it Atomic Spectra and Radiative Transitions},  Springer-Verlag (Berlin, 1996).
\bibitem{RyabtsevTretyakov2002} I. I. Ryabtsev and D. B. Tretyakov, "The Break of L-S Coupling and the Double Stark Resonance in the Spectrum of 36P-37P Two-Photon Transition in Rydberg Atoms of Sodium", JETP {\bf 94}, 677-684 (2002). 
\bibitem{BerkelandBoshier2002} D. J. Berkeland, and M. G. Boshier, "Destabilization of dark states and optical spectroscopy in Zeeman-degenerate atomic systems", Phys. Rev. A {\bf 65}, 033413 (2002).
\bibitem{Nienhuis1996} G. Nienhuis, "Natural basis of magnetic substates for a radaitive transition with arbitrary polarization", Opt. Commun. {\bf 59}, 353 (1996).
\bibitem{MilnerPrior1998} V. Milner, and Y. Prior, "Multilevel dark states: coherent population trapping with elliptically polarized incoherent light", Phys. Rev. Lett. {\bf 80}, 940 (1998).
\bibitem{Prudnikov2004} O.N. Prudnikov,  A.V.  Taichenachev,  A.M. Tumaikin,  V. I. Yudin, and G. Nienhuis,  "Basis of polarization-dressed states of an atom in an elliptically polarized resonant field", Sov. Phys. JETP { \bf99}, 1137-49 (2004).
\bibitem{ZlatanovVitanv2020}   K. N. Zlatanov, G. S. Vasilev, and N. V. Vitanov, "Morris-Shore transformation for nondegenerate systems", Phys. Rev.  A {\bf 102}, 063113 (2020).
\bibitem{LiGallagher2003} W. Li, I. Mourachko, M. W. Noel, and T. F. Gallagher, "Millimeter-wave spectroscopy of cold Rb Rydberg atoms in a magneto-optical trap: Quantum defects of the ns, np, and nd series", Phys. Rev. A {\bf 67}, 052502 (2003).\bibitem{SibalicAdamsWeatherhill2017} N. Sibali\'c, J.D. Pritchard,  C.S. Adams and K.J. Weatherill, "ARC: An open-source library for calculating properties of alkali Rydberg atoms", Comp. Phys. Comm. {\bf 220}, 319-331 (2017).
\bibitem{KisVitanov2003} Z. Kis, A. Karpati, B. W. Shore, and N. V. Vitanov, "Stimulated Raman adiabatic passage among degenerate-level manifolds", Phys. Rev. A {\bf 70}, 053405 (2004).
\bibitem{KirovaEkers2017} T. Kirova, A. Cinins, D. K. Efimov, M. Bruvelis, K. Miculis, N. N. Bezuglov, M. Auzinsh, I. I. Ryabtsev, and A. Ekers,  "Hyperfine interaction in the Autler-Townes effect: The formation of bright, dark, and chameleon states", Phys. Rev. A {\bf 96}, 043421 (2017).
\end{thebibliography}
\end{document}